\documentclass[12pt,aps,showpacs,nofootinbib,showkeys]{revtex4}
\usepackage{amsmath}    
\usepackage{graphicx}   
\usepackage{color}      
\usepackage{epstopdf}
\usepackage{setspace}
\usepackage{amsfonts}
\usepackage{ulem}
\usepackage[dvipsnames]{xcolor}
\usepackage{tikz}
 \usepackage{pgfplots}
\usepackage{feynmp-auto}
\usepackage{tikz}
\usepackage{graphicx}
\usepackage{amssymb}
\usepackage{amsmath}
\usepackage{graphics}
\usepackage{dcolumn}
\usepackage{bm}
\usepackage{color}
\usepackage{epstopdf}
\usepackage{lipsum}
\usepackage{ulem}
\usepackage{braket}
\usepackage{verbatim}
\usepackage{hyperref} 

\begin{document}


\newcommand{\blue}[1]{\textcolor{blue}{#1}}
\newcommand{\new}{\blue}
\newcommand{\green}[1]{\textcolor{green}{#1}}
\newcommand{\modif}{\green}
\newcommand{\red}[1]{\textcolor{red}{#1}}
\newcommand{\attention}{\red}


\title{Mimicking Negative Mass Properties}

\author{S. D.  Campos}\email{sergiodc@ufscar.br}
	
\address{Applied Mathematics Laboratory - CCTS/DFQM,
Federal University of S\~ao Carlos, Sorocaba, CEP 18052780, Brazil}

\begin{abstract}
In the present work, one analyzes two systems trying to obtain physical conditions where some properties attributed to negative mass can be mimicked by positive mass particles. The first one is the well-known 1/2-spin system described by the Dirac equation in the presence of an external electromagnetic field. Assuming some physical restrictions, one obtains that the use of $e\rightarrow-e$ can lead to the same results as using $m\rightarrow-m$. In particular, for a null dielectric function, it is possible to obtain a negative mass behavior from a positive mass system composed of negatively charged particles. The second system is based on the de Broglie matter wave. The dispersion relation of such a wave can be negative (real or imaginary valued) if one assumes an imaginary wavenumber. The consequence is the emergence of a negative refractive index for positive mass particles. However, this behavior is generally attributed to a negative mass system.
\end{abstract}

\pacs{11.55.Fv, 05.30.Fk, 71.70.Ej}

\keywords{negative mass; Dirac equation; dispersion relation; spin-orbit}




\maketitle
\section{Introduction}

For some intriguing reason, unicorns do not exist, even though the laws of Nature do not explicitly prohibit their existence. The general belief is that these laws allow for any phenomenon unless there is a specific constraint or limitation explicitly preventing it. The laws of Nature are inclusive rather than restrictive unless stated otherwise. Hence, it is theoretically possible to create one of these fantastic creatures. Of course, it is not a true-born unicorn since it was built following some recipe (not necessarily the \textit{best one}) and obeying some (not necessarily \textit{all}) rules displayed on the table.

In physics, there is no fundamental law of classical or quantum mechanics prohibiting the existence of negative mass (see Ref. \cite{SDC.2023} and references therein). However, the lack of experimental evidence in the present day unfortunately suggests that negative mass does not exist. Indeed, the main assumption in \cite{SDC.2023} is that negative mass particles have imaginary velocity, which prohibits their direct experimental determination. However, the transformation $v\rightarrow iv$ provides positive kinetic energy, certainly contributing to the total energy of the system.  

One of the first alleged experimental evidence of a system presenting physical properties related to negative effective mass is from 1959 \cite{G.C.Dousmanis.R.C.DuncanJr.J.J.Thomas.R.C.Williams.PRL.1(11)404.1958}. Using cyclotron resonance, Dousmanis and collaborators suggested the detection of negative mass carriers in Ge crystals at 4 K. In 2017, more than 60 years later, some physical effects expected for negative effective mass were generated in a spin-orbit coupled Bose-Einstein Condensate \cite{M.A.Khamehchi.K.Hossain.M.E.Mossman.Y.Zhang.Th.Busch.M.M.Forbes.P.Engels.PRL118.155301.2017}. Specifically, Khamehchi and collaborators suggest that a modified dispersion relation and a negative effective mass may explain their results as well as other related \cite{J.P.Ronzheimer.M.Schreiber.S.Braun.S.S.Hodgman.S.Langer.I.P.McCulloch.F.Heidrich.Meisner.I.Bloch.U.Schneider.PRL110.205301.2013,
A.Reinhard.J.Riou.L.A.Zundel.D.S.Weiss.S.Li.A.M.Rey.R.Hipolito.PRL110.033001.2013,
K.Henderson.H.Kelkar.B.Gutierrez.Medina.T.C.Li.M.G.Raizen.PRL96.150401.2006,
Th.Anker.M.Albiez.R.Gati.S.Hunsmann.B.Eiermann.A.Trombettoni.M.K.Oberthaler.PRL94.020403.2005,
B.Eiermann.P.Treutlein.Th.Anker.M.Albiez.M.Taglieber.K.P.Marzlin.M.K.Oberthaler.Phys.Rev.Lett91.060402.2003}.

Here, in the first step, the system studied describes the Dirac equation in the presence of an external electromagnetic field, showing that $m\rightarrow-m$ is not everywhere equivalent to $e\rightarrow -e$. Indeed, both transformations describe the same system only when some physical conditions, involving the scalar electric field or the dielectric function, are assumed to be valid. An interesting case emerges when the dielectric function is null for some frequencies. In this case, a positively charged system with negative mass has the same physical behavior as a negatively charged system with positive mass. Similar behavior can be viewed in some metamaterials where a negative mass behavior emerges from the plasma oscillations of a free electron gas \cite{E.Bormashenko.I.Legchenkova.Materials.Basel.13.8.1890.2020,E.Bormashenko.I.Legchenkova.M.Frenkel.Materials.Basel.13.16.3512.2020}. There are several applications for metamaterials with properties not found in usual materials. In particular, some metamaterials allow the emergence of a negative refractive index \cite{M.A.Khamehchi.K.Hossain.M.E.Mossman.Y.Zhang.Th.Busch.M.M.Forbes.P.Engels.PRL118.155301.2017,J.A.Dionne.Optics.Express.16.23.19001.2008,J.Valentine.Nature.455.376.2008}, connected with the next case studied here.

The second step presents the de Broglie matter wave dispersion relation.  Rather than resorting to a negative mass for achieving a negative dispersion relation, it is obtained by employing an imaginary wavenumber, indicating the wave decaying within the medium. The resulting group velocity is imaginary, expressing the presence of a non-conservative force. Depending on the frequency, a negative imaginary dispersion relation can also arise, with the corresponding imaginary group velocity pointing out a non-restorative force affecting the system. In any case, the negative real or negative imaginary dispersion relation may be viewed as being due to a negative mass \cite{Y.Liu.G.P.Wang.J.B.Pendry.S.Zhang.Light.Sci.Appl.11.276.2022,T.Suzuki.M.Sekiya.T.Sato.Y.Takebayashi.Opt.Express.26.8314.2018,M.A.Khamehchi.K.Hossain.M.E.Mossman.Y.Zhang.Th.Busch.M.M.Forbes.P.Engels.PRL118.155301.2017,D.Forcella.C.Prada.R.Carminati.Phys.Rev.Lett.118.134301.2017,S.A.Ramakrishna.Rep.Prog.Phys.68.449.2005,K.Y.Bliokh.Y.P.Bliokh.Sov.Phys.Uspekhi.47.393.2004,Y.Ben.Aryeh.J.Mod.Opt.52.1871.2005,J.B.Pendry.Phys.Rev.Lett.85.3966.2000}.

This paper is organized as follows. In section \ref{sec:dirac}, one shows that the use of a negative mass in the Dirac equation (free-particle) is equivalent to a time-reversal procedure. However, in the presence of an electromagnetic 4-vector, it is not generally true as shall be seen in Section \ref{sec:electro}.  Section \ref{sec:disprel} presents some considerations about the negative dispersion relation. Discussion and remarks are left for the final section \ref{sec:discussion}.

\section{Negative Mass and Charge in the Dirac Equation}\label{sec:dirac}

The use of $-m$ instead of $m$ in the Dirac equation (DE) seems to be a simple mathematical replacement. However, this alteration could carry significant physical ramifications since the interaction properties of negative mass are quite different from positive ones (see Ref. \cite{SDC.2023} and references therein). 

In this first case, the DE is used to describe $1/2$-spin particles in the absence of a potential, being written as 
\begin{eqnarray}\label{eq:dirac1}
  \mathcal{H}\psi(t,\mathbf{r})= i\hbar\frac{\partial\psi(\mathbf{r},t)}{\partial t},
\end{eqnarray}
\noindent where $\mathbf{r}=(x^1,x^2,x^3)$ and
\begin{eqnarray}
    \mathcal{H}=\mathbf{\alpha}\cdot\bigl(-i\hbar\mathbf{\nabla}\bigr)-\alpha_4 mc^2
\end{eqnarray} 
\noindent for $\mathbf{\alpha}=(\alpha_1,\alpha_2,\alpha_3)$. For a matter of internal consistency, $\alpha_i$'s obey the following known relations ($i,j=1,2,3$)
\begin{eqnarray}\label{eq:set_1}
    \alpha_4^2=\mathcal{I}, ~~ \{\alpha_4,\alpha_i\}=0, ~~\{\alpha_i,\alpha_j\}=2\delta_{ij}\mathcal{I},
\end{eqnarray}
\noindent where $\{a,b\}\equiv  ab+ba$ is the anticommmutator and $\mathcal{I}$ is the $4\times 4$ identity matrix. 
Using $\gamma$ matrices, one can rewrite DE (\ref{eq:dirac1}) in the compact form \cite{mandl.shawn}
\begin{eqnarray}\label{eq:dirac_xx}
    (i\hbar\gamma^{\mu}\partial_{\mu}-mc)\psi(\mathbf{r},t)=0.
\end{eqnarray}

From the mathematical point of view, replacing $m$ with $-m$ does not alter the solutions of DE: it still yields positive energy solutions for electrons and negative energy solutions for positrons. Then, the equation
\begin{eqnarray}\label{eq:dirac_xxxx}
    (i\hbar\gamma^{\mu}\partial_{\mu}\pm mc)\psi(\mathbf{r},t)=0,
\end{eqnarray}
\noindent where $+$ and $-$ signs stand for positive and negative mass, respectively, has the same set of solutions (interpreted according to the mass sign). For the sake of simplicity, one looks here just for the stationary solutions of \eqref{eq:dirac_xxxx}, written as
\begin{eqnarray}\label{eq:stationary}
    \psi(\mathbf{r},t)=\chi(\mathbf{r})e^{- i(\mathcal{E}/\hbar)t},
\end{eqnarray}
\noindent where $\mathcal{E}$ is the energy that can be positive (particles) or negative (antiparticles). The two-component of the time-independent spinor field $\chi(\mathbf{r})$ is given by
\begin{eqnarray}
    \chi(\mathbf{r})=\begin{pmatrix}
\phi(\mathbf{r}) \\
\eta(\mathbf{r}) \\
\end{pmatrix}.
\end{eqnarray}

Nevertheless, is there another parameter within the DE that, when replaced, could yield equivalent physical interpretations for the solutions obtained when using $m\rightarrow-m$? The answer is yes.

It is straightforward to observe that the transformation $t\rightarrow-t$ leads to the same set of solutions if one assumes $+\mathcal{E}$ for the particle and $-\mathcal{E}$ for the antiparticle. Consequently, from a dynamics standpoint, replacing $m\rightarrow -m$ in DE is \textit{equivalent} to using $t\rightarrow -t$ in the same differential equation, taking into account the convenient definition of particles and antiparticles. 

As shall be seen, the presence of an electromagnetic field introduces some interesting physical results when one uses $m\rightarrow-m$ or $e\rightarrow-e$.

\section{Dirac Equation in the Presence of an External Electromagnetic Field}\label{sec:electro}

\subsection{Basic Picture}

For the sake of simplicity, let us assume that the vector $\mathbf{A}=\mathbf{A}(\mathbf{r})$ and the scalar $\phi=\phi(\mathbf{r})$ potentials are time-independent, and associated with the magnetic and electric components of the external electromagnetic field by the relations \cite{jackson}
\begin{eqnarray}
\nonumber    \mathbf{B}=\nabla\times\mathbf{A}, ~~ \mathbf{E}=-\nabla\phi,
\end{eqnarray}
\noindent where $\mathbf{B}=\mathbf{B}(\mathbf{r})$ and $\mathbf{E}=\mathbf{E}(\mathbf{r})$.

The Dirac equation for a particle with charge $q$ subject to an external electromagnetic field has the following Hamiltonian \cite{bransden.joachin} 
\begin{eqnarray} \label{eq:hamilton3}
    \mathcal{H}=\frac{1}{2m}\bigl[\mathbf{p}-q\mathbf{A}\bigr]^2+q\phi-\frac{q\hbar}{2m}\bigl[\mathbf{\sigma}\cdot\mathbf{B}\bigr],
\end{eqnarray}
\noindent where the linear momentum $\mathbf{p}$  is defined as $\mathbf{p}=-i\hbar\mathbf{\nabla}$ and the term $e\hbar/2m\bigl(\mathbf{\sigma}\cdot\mathbf{B}\bigr)$ corresponds to the interaction term relating the spin and the magnetic field $\mathbf{B}$, and, for the sake of simplicity, hereafter $m=m_e=m_p$.  
The spin and magnetic moment operators are related by
\begin{eqnarray}\label{eq:spin_moment}
    \mathbf{\mu}_q=\frac{q\hbar}{2m}\mathbf{S},
\end{eqnarray}
\noindent and for the electron, it is usual to represent the above equality as
\begin{eqnarray}\label{eq:spin_moment1}
    \mathbf{\mu}_e=-\mu_B\mathbf{\sigma}=-\frac{e\hbar}{2m}\mathbf{S}
\end{eqnarray}
\noindent to denote that $\mathbf{\mu}_e$ and $\mathbf{S}=(\hbar/2)\mathbf{\sigma}$ are antiparallel vectors \cite{M.Vogel.W.Quint.Phys.Rep.490.1.2010}. 
Moreover, $\mu_B$ is the Bohr magneton, and $\sigma=(\sigma_x,\sigma_y,\sigma_z)$ are the usual Pauli spin matrices.

In the case of electrons $-e$ and positrons $+e$, replacing \eqref{eq:hamilton3} in \eqref{eq:dirac_xx} should be followed by the Dirac prescription \cite{P.A.M.Dirac.Proc.Roy.Soc.Lond.Ser.A.133.80.1931,Bethe.Salpeter_book}: transformation $\mathcal{E}\rightarrow-\mathcal{E}$ is also followed by changing the sign of $\mathbf{p}$, $e\mathbf{A}$, $e\phi$, and the interaction term. This procedure ensures the solutions for negative $\mathcal{E}$ are only the opposite of the solutions for positive $\mathcal{E}$.  The results for positive and negative $\mathcal{E}$ are the well-known Pauli equations (for electrons and positrons)
\begin{eqnarray}\label{eq:pauli_1}
\mathcal{E}_\pm\chi_\pm (\mathbf{r})&=&\pm\left[\frac{1}{2m}\bigl(\mathbf{p}+ e\mathbf{A}\bigr)^2- e\phi +\frac{e\hbar}{2m}\bigl(\mathbf{\sigma}\cdot\mathbf{B}\bigr)\right]\chi_\pm(\mathbf{r}).
\end{eqnarray}

Notice, as expected, that $\mathcal{E}_-=-\mathcal{E}_+$. 
On the other hand, the transformation $t\rightarrow-t$ furnishes the same result as above only if the Dirac prescription is used. It seems reasonable since the transformation $t\rightarrow -t$ is equivalent to $\mathcal{E}\rightarrow -\mathcal{E}$ in the stationary solutions. One writes for electrons and positrons (note that $t\rightarrow-t$ implies $dt\rightarrow -dt$)
\begin{eqnarray}\label{eq:pauli_003}
\mathcal{E}_\pm\chi_\pm (\mathbf{r})&=&\pm \left[\frac{1}{2m}\bigr(\mathbf{p}+ e\mathbf{A}\bigl)^2- e\phi + \frac{e\hbar}{2m}\bigl(\mathbf{\sigma}\cdot\mathbf{B}\bigr)\right]\chi_\pm(\mathbf{r}).
\end{eqnarray}

Comparing \eqref{eq:pauli_1} and \eqref{eq:pauli_003}, one observes the well-known result that the time and charge inversion represent the same physical situation for the Dirac equation in the presence of an external electromagnetic field.

The interest here is to study the effects of the transformation $m\rightarrow-m$ in the Dirac equation using the above Hamiltonian and looking for stationary solutions \eqref{eq:stationary}. Let us start by using the stationary solution \eqref{eq:stationary} and replacing $m\rightarrow -m$  into DE. The results can be exhibited in a single equation for electrons and positrons 
\begin{eqnarray}\label{eq:pauli_002}
\mathcal{E}_\pm\chi_\pm(\mathbf{r})=\mp\left[ \frac{1}{2m}\bigl(\mathbf{p}+ e\mathbf{A}\bigr)^2 + e\phi + \frac{e\hbar}{2m}\bigl(\mathbf{\sigma}\cdot\mathbf{B}\bigr)\right]\chi_\pm(\mathbf{r}).
\end{eqnarray}

Comparing \eqref{eq:pauli_002} and \eqref{eq:pauli_1}, one observes that in the DE equation for the Hamiltonian \eqref{eq:hamilton3}, the transformations $e\rightarrow-e$ and $m\rightarrow -m$ do not lead to the same equations. However, these transformations represent the same physical situation if the following condition is satisfied 
\begin{eqnarray}\label{eq:sot}
    e\phi\chi(\mathbf{r})=0.
\end{eqnarray}

The simplest way to accomplish the above condition is assuming the static charge distribution $\rho=\rho(\mathbf{r})$ in a given volume $V$ is null 
\begin{eqnarray}
    \phi(\mathbf{r})=\frac{1}{4\pi\epsilon_0}\int \frac{\rho}{|\mathbf{r}-\mathbf{r}'|}dV=0.
\end{eqnarray}
\noindent which occurs for a zero charge density within the homogeneous volume $V$ and a transverse electric field. In general, result \eqref{eq:sot} implies
\begin{eqnarray}
    \nabla\cdot \mathbf{E}=-\nabla^2\phi=0.
\end{eqnarray}

However, there is another way to obtain \eqref{eq:sot} without assuming $\nabla\cdot \mathbf{E}=0$. Notice the Gauss law can be written by using the displacement field $\mathbf{D}=\mathbf{D}(\mathbf{r})=\epsilon(\omega)\mathbf{E}+4\pi\mathbf{P}$ as \cite{jackson}
\begin{eqnarray}
   \nabla\cdot \mathbf{D}=\rho,
\end{eqnarray}
\noindent where $\epsilon(\omega)$ is the dielectric function that depends on the frequency $\omega$, and the polarization field $\mathbf{P}$ is associated with the induced charge density. For a linear homogeneous isotropic material, the polarization is null. Hence, one writes 
\begin{eqnarray}\label{eq:prod}
    \epsilon(\omega)\bigr[\nabla\cdot \mathbf{E}\bigl]=0.
\end{eqnarray}

The product shown in result \eqref{eq:prod} is null $\nabla\cdot \mathbf{E}=0$ or for $\epsilon(\omega)=0$, for some frequencies. Assuming $\epsilon(\omega)=0$, for some frequencies, it implies the existence of a non-null longitudinal contribution for the electric field allowing the plasma to have a charge fluctuation, where the resulting plasma oscillation is the so-called plasmon \cite{Stefan.A.Maier.Plasmonics.Fundamentals.and.Applications.Springer.2007}. 

The above assumption on the dielectric function implies that the plasmon emerging in a plasma composed of negatively charged particles (with positive mass) behaves as a plasmon emerging in a positively charged plasma (with negative mass). An effect similar to this one was already observed in metamaterials \cite{E.Bormashenko.I.Legchenkova.Materials.Basel.13.8.1890.2020,E.Bormashenko.I.Legchenkova.M.Frenkel.Materials.Basel.13.16.3512.2020}, where plasma oscillations of a free electron gas induces a negative effective mass in the system due to the vibrations of a metallic particle with frequency $\omega$, close to the plasma oscillation frequency \cite{E.Bormashenko.I.Legchenkova.Materials.Basel.13.8.1890.2020}.

\section{Dispersion Relation}\label{sec:disprel}

Considering massive particles, the de Broglie matter waves can be obtained by considering the usual relativistic energy relation 
\begin{eqnarray}
    (\hbar \omega)^2=(pc)^2+\bigl(mc^2\bigr)^2,
\end{eqnarray}
\noindent where $\mathbf{p}=\hbar\mathbf{k}$, $\mathbf{k}$ is the wavevector ($|\mathbf{k}|=k$), and $\omega=\omega(k)$ is the angular frequency (also known as dispersion relation). Solving the above equation for $\omega$, one obtains the well-known relativistic dispersion relation for matter waves
\begin{eqnarray}\label{eq:omega1}
    \omega= \frac{m c^2}{\hbar}\sqrt{1+\left(\frac{\hbar k}{m c}\right)^2}.
\end{eqnarray}

Observe that \eqref{eq:omega1} can assume negative values by considering 
$m<0$ and a convenient range of frequency. However, the main purpose here is to present physical conditions where $\omega$ can be negative without assuming \textit{a priori} negative mass since one seeks physical situations where a positive mass can mimic the expected behavior of a negative mass.

Indeed, a similar result can be achieved by assuming that 
$k$ is an imaginary number representing an exponential decay away from the source. Thus, the wavenumber is (almost) locally defined near the source. Indeed, this condition occurs in the so-called evanescent wave and indicates it does not propagate, representing a local vibration with fast decay away from the source \cite{V.Giurgiutiu.M.F.Haider.Materials.Basel.12.2.269.2019}. Then, one writes the imaginary wavenumber as
\begin{eqnarray}\label{eq:ansatz}
    k=i\delta,
\end{eqnarray}
\noindent where $\delta$ corresponds to the decaying of the wave energy through the medium. Then, inserting the assumption \eqref{eq:ansatz} into equation \eqref{eq:omega1}, one obtains 
\begin{eqnarray}\label{eq:omega2}
    \omega&=& -\frac{m_0 c^2}{\hbar}\sqrt{1-\left(\frac{\hbar \delta}{m_0 c}\right)^2}.
\end{eqnarray}

Notice the emergence of the negative sign within the square root imposes some care in the study of the dispersion relation \eqref{eq:omega2}. Indeed, the above dispersion relation is negative if
\begin{eqnarray}\label{eq:inel1}
    -\frac{m_0 c}{\hbar}< \delta<\frac{m_0 c}{\hbar},
\end{eqnarray}
\noindent while it is imaginary for $\delta>|m_0c/\hbar|$. 

It is important to stress that replacing $m\rightarrow -m$ into \eqref{eq:omega1} is equivalent to using the imaginary wavenumber only if the bound \eqref{eq:inel1} is satisfied.  Thus, the dispersion relation \eqref{eq:omega2}, in this simple case, is not due to the mass sign but to the introduction of an imaginary wavenumber, that represents the introduction of a non-conservative force. Therefore, the introduction of the imaginary wavenumber allows a positive mass to have a negative dispersion relation. 

A similar behavior is expected for light in some metamaterials, where the medium bends light to a negative angle relative to the surface normal, i.e. the incident and refracted waves lie on the same side of the surface normal \cite{Y.Liu.G.P.Wang.J.B.Pendry.S.Zhang.Light.Sci.Appl.11.276.2022}. This behavior is exposed in \cite{M.A.Khamehchi.K.Hossain.M.E.Mossman.Y.Zhang.Th.Busch.M.M.Forbes.P.Engels.PRL118.155301.2017}, where the negative dispersion is argued to indicate a negative refraction index. There is a growing interest in materials with a negative refraction index \cite{Y.Liu.G.P.Wang.J.B.Pendry.S.Zhang.Light.Sci.Appl.11.276.2022,T.Suzuki.M.Sekiya.T.Sato.Y.Takebayashi.Opt.Express.26.8314.2018,M.A.Khamehchi.K.Hossain.M.E.Mossman.Y.Zhang.Th.Busch.M.M.Forbes.P.Engels.PRL118.155301.2017,D.Forcella.C.Prada.R.Carminati.Phys.Rev.Lett.118.134301.2017,S.A.Ramakrishna.Rep.Prog.Phys.68.449.2005,K.Y.Bliokh.Y.P.Bliokh.Sov.Phys.Uspekhi.47.393.2004,Y.Ben.Aryeh.J.Mod.Opt.52.1871.2005,J.B.Pendry.Phys.Rev.Lett.85.3966.2000}. Causality is invoked to argue that dissipation is a necessary condition for a negative refraction index \cite{M.I.Stockman.Phys.Rev.Lett.98.177404.2007}. In the case of matter waves presented here, the introduction of an imaginary wavenumber means the introduction of a non-restorative force (respecting causality).

On the other hand, some values of $\delta$ can also allow an imaginary dispersion relation, representing the absorption of the wave by the medium. Here, the imaginary dispersion relation \eqref{eq:omega2} is negative, indicating the decaying amplitude in the direction of the wave motion. This result agrees with the introduction of a dissipative force.

For the case of a negative real dispersion relation, the group velocity obtained from \eqref{eq:omega2} is imaginary ($dk=id\delta$) and opposite to the dispersion relation, being given by
\begin{eqnarray}\label{eq:domega2}
    \frac{d\omega}{d\delta}=\frac{i\hbar \delta}{m_0\sqrt{1-\left(\frac{\hbar \delta}{m_0}\right)^2}}.
\end{eqnarray}

The above result is typical of evanescent modes \cite{J.W.Ryu.N.Myoung.H.C.Park.Scientific.Reports.7.8746.2017},  indicating values of frequency for which the wave does not propagate. Notice the energy related to this velocity is negative $E\sim (id\omega(\delta)/d\delta)^2$ since the mass is positive.

On the other hand, considering a negative imaginary dispersion relation, the group velocity
is negative real, 
\begin{eqnarray}\label{eq:domega02}
    \frac{d\omega}{d\delta}=-\frac{\hbar \delta}{m_0\sqrt{1-\left(\frac{\hbar \delta}{m_0}\right)^2}},
\end{eqnarray}
\noindent which can be interpreted as a negative refraction index \cite{V.Veselago.L.Braginsky.V.Shklover.C.Hafner.J.Comput.Theor.Nanosci.3.1.2006}. 

It is important to stress that the second derivative of the frequency \eqref{eq:omega2} should be positive to ensure the minimum value of the potential is achieved. This condition holds if
\begin{eqnarray}\label{eq:inel2}
    -\frac{m_0}{\hbar}<\delta<\frac{m_0}{\hbar},
\end{eqnarray}
\noindent implying the dispersion relation is negative real while the group velocity is imaginary since $(-m_0/\hbar,m_0/\hbar)\in(-m_0 c/\hbar,m_0 c/\hbar)$. Nevertheless, if this condition \eqref{eq:inel2} is released, then the dispersion relations can also be negative imaginary, whereas the group velocity is negative real. 

Therefore, if one assumes $k=i\delta$, one obtains two distinct behaviors for the dispersion relation: (a) a negative real dispersion relation and an imaginary group velocity or (b) a negative imaginary dispersion relation and a negative real group velocity. Both cases depend on the range of $\delta$, i.e. the frequency of the system.  However, whether the dispersion relation real or imaginary valued function, the negative refraction index seems to be a natural consequence of that.



\section{Discussion and Remarks}\label{sec:discussion}

The present work focuses on studying the physical conditions where a positive mass can mimic the behavior of a negative one. In the first case of study, one uses $m\rightarrow -m$ in the Dirac equation in the presence of an external electromagnetic field. 

The equivalence between $m\rightarrow -m$ and $e\rightarrow-e$ can be achieved under the assumptions that (a) the scalar field $\phi$ is null or (b) the dielectric function is null. The latter furnishes the possibility of a negative mass behavior emerging, for example, from a plasma negatively charged with a positive mass \cite{E.Bormashenko.I.Legchenkova.Materials.Basel.13.8.1890.2020,E.Bormashenko.I.Legchenkova.M.Frenkel.Materials.Basel.13.16.3512.2020}. This possibility may allow the construction of metamaterials with specific properties and, in particular, with a negative refraction index, which is the subject of the second system studied here.

In the de Broglie matter wave the emergence of a negative dispersion relation can be achieved without assuming a negative mass. Indeed, it can be done by assuming an imaginary wavenumber representing the decaying of the wave energy through the medium. This assumption, for some range of frequencies, leads to (a) a negative dispersion relation or (b) an imaginary negative dispersion relation. The condition (a) results in an imaginary negative group velocity while (b) the group velocity is negative real. In both cases, the consequence is the emergence of a negative refraction index, which may be attributed to a negative mass \cite{Y.Liu.G.P.Wang.J.B.Pendry.S.Zhang.Light.Sci.Appl.11.276.2022,T.Suzuki.M.Sekiya.T.Sato.Y.Takebayashi.Opt.Express.26.8314.2018,M.A.Khamehchi.K.Hossain.M.E.Mossman.Y.Zhang.Th.Busch.M.M.Forbes.P.Engels.PRL118.155301.2017,D.Forcella.C.Prada.R.Carminati.Phys.Rev.Lett.118.134301.2017,S.A.Ramakrishna.Rep.Prog.Phys.68.449.2005,K.Y.Bliokh.Y.P.Bliokh.Sov.Phys.Uspekhi.47.393.2004,Y.Ben.Aryeh.J.Mod.Opt.52.1871.2005,J.B.Pendry.Phys.Rev.Lett.85.3966.2000}.

As a last comment, it should be pointed out that the Klein-Gordon equation is a relativistic second-order differential equation that correctly describes spinless composite particles, written as
\begin{eqnarray}
    \frac{1}{c^2}\frac{\partial^2 \psi(\mathbf{r},t)}{\partial t^2}=\nabla^2\psi(\mathbf{r},t)-\frac{m^2c^2}{\hbar^2}\psi(\mathbf{r},t).
\end{eqnarray}

The use of $m\rightarrow -m$ or $t\rightarrow -t$ is indifferent, leading to the same set of solutions for $\psi(\mathbf{r},t)$. It indicates that in the Klein-Gordon equation, there is no physical difference between particles with positive or negative mass. It corroborates to the usual understanding that the spinless particles are also their antiparticles. 



\section*{Acknowledgments}

The author thanks to UFSCar.


\end{document}